# Bias Variance Tradeoff in Analysis of Online Controlled Experiments

Ali Mahmoudzadeh, Sophia Liu, Sol Sadeghi, Paul Luo Li, Somit Gupta
Microsoft
1 Microsoft Way
Redmond, WA USA
{Ali.Mahmoudzadeh; Soheil.Sadeghi; Paul.Li; Somit.Gupta}@microsoft.com, SLiu@Netflix.com

## ABSTRACT

Many organizations utilize large-scale online controlled experiments (OCEs) to accelerate innovation. Having high statistical power to detect small differences between control and treatment accurately is critical, as even small changes in key metrics can be worth millions of dollars or indicate user dissatisfaction for a very large number of users. For large-scale OCE, the duration is typically short (e.g. two weeks) to expedite changes and improvements to the product. In this paper, we examine two common approaches for analyzing usage data collected from users within the time window of an experiment, which can differ in accuracy and power. The 'open' approach includes all relevant usage data from all active users for the entire duration of the experiment. The 'bounded' approach includes data from a fixed period of observation for each user (e.g. seven days after exposure) after the first time a user became active in the experiment window. We study the effect of these two approaches on double averaged metrics by mathematically modeling their properties under various assumptions and testing them in real world for a class of metrics for service (Bing) and client (Edge) experiments. While all users of an online service can be exposed to a treatment right after the change is deployed – thus there is a fixed target population, users of a client must upgrade to the right version to be exposed to the treatment – thus there is an evolving target population. We observe that in case of evolving target population (client experiments) the bounded approach may have reduced bias (i.e. improved accuracy) but at the cost of increased variance (i.e. low power). In the case of fixed target population (service experiments), the open approach has reduced bias and also low variance. Practically, for client experiments, the bias in open approach is small and does not change decision outcome. Based on the findings, we make use of the open approach in most cases. In addition, we also suggest areas of future work for researchers.

## 1 Introduction

Online controlled experiments (e.g., A/B tests) are rapidly becoming a gold standard for evaluating improvements to web sites, mobile applications, desktop applications, services, and operating systems[19]. Even small product changes can result in millions of dollars in revenue [21].

In typical A/B tests, users are randomly assigned to treatment or control groups. Over the predetermined experimental period, usage data is collected from users and processed/aggregated into metrics. Finally, statistical tests are performed on the metrics to detect differences between treatment and control groups (that are unlikely to happen due to chance). The key benefit of a randomized controlled experiment is its ability to establish causal impact of a change[15,26]. Randomization ensures that other factors that can impact outcomes are balanced between control and treatment groups, such that the change made to the product is the only explanation for the differences between treatment and control, outside of random noise in the metrics.

### 1.1 Related Work

Many organizations utilize large-scale digital OCEs to accelerate innovation. Having high statistical power to detect small differences between control and treatment accurately is essential to success, as even small changes in key metrics can be worth millions of dollars or indicate user dissatisfaction for a very large number of users [18,21]. For large-scale OCEs, the duration is typically short (e.g. two weeks) to expedite changes and improvements to the product. There are two common approaches for analyzing usage data collected from users within the time window of an experiment, which can differ in accuracy and power. The 'open' approach includes all relevant usage data from all active users for the entire duration of the experiment. Thus, we include all users who were active, but the period of actual observation may vary by user based on when the user first became active in the experiment window. The 'bounded' approach includes data from a fixed period of observation for each user (e.g. seven days after exposure) after the first time a user became active in the experiment window. If a user does not have the needed amount of observation days (e.g. a user arriving on the last day of the experiment window), then the user is not included in the analysis, neither is the data from days beyond the observation window (e.g. no data from the eighth day). Relative to open approach, the bounded approach is less sensitive to when a user first became active in the experiment window but it includes fewer users.

Different experimentation teams have adopted one of bounded and open data inclusion analysis approaches. However, existing literature has not studied the trade-offs between bounded and open data inclusion analysis approaches with rigor and with real-world data. In this paper, we start such comparison by studying the effect on doubly averaged metrics. We make the following contributions:
- Mathematically model accuracy and power of bounded and open approaches under various assumptions for client and service experiments
- Empirically evaluate both approaches using real-world data, through both simulations and real large-scale OCEs
- Share our preferred approach to use in most cases

We find that for time-invariant homogeneous treatment effects (or where heterogeneity is small), for the class of doubly averaged metrics, both approaches will produce an unbiased (i.e. accurate) estimate of the treatment effect. We show the open approach is well suited for service experiments with low bias (i.e. accurate) and low variance (high statistical power). We also show that the open approach can have biased estimates in the presence of weekday/weekend effects for client experiments. While the bounded approach is without such bias, it sacrifices substantial statistical power due to increase in variance. In practice the bias is small and, we have experienced, the ship decision generally did not change due to small bias. Therefore, in general, we conclude that the open approach is the practical choice for most experimentation scenarios.

The rest of the paper is organized as follows. We start by introducing a mathematical framework for studying open and bounded analysis in Section 2. In Section 3, we mathematically model the bias and variance estimates for the bounded and open approaches under various assumptions. In Section 4 and Section 5, we compare the two approaches, using simulations and data from real-world experiments. In Section 6, we conclude by summarizing recommendations to practitioners and making suggestions for future research.

## 2 Background

### 2.1 Related Work

There has been a lot of research in the area of improving statistical power for experimentation metrics [4,5,9,16,24,28] under different conditions and using different types of covariates and metric design. There is also a growing area of research on experiment design, metric design, and good experimentation system architecture for trustworthy analysis at scale [3,6–8,10,11,13,17,18,22,25]. Some of these papers also study the trade-off between bias and variance [10,23] in different settings. The subject of validity of a controlled experiment has also been studied in other fields of research like psychology and public policy [1,2]. To the best of our knowledge this is the first paper that discusses the bias variance tradeoff between the open and bounded data inclusion analysis approaches.

### 2.2 Framework

Our framework relies on the following two assumptions:

**Assumption 1 (Stable Unit Treatment Value Assumption):**

One user's outcome is unaffected by other users' treatment assignments. This is a standard assumption used in most A/B tests [14].

**Assumption 2 (Incremental Experiment Assumption):**

We assume that for each user $u$, the days a user is active is independent of the treatment assignment. It implies that our experiment is incremental such that it does not change the behavior of users' active days (or visits in general). Based on our experience, this assumption is mild since user visit patterns rarely change during the experiment [18]. However, it can bring limitations to the model if the treatment causes dramatic change in user visit patterns. We emphasize that before using our framework, we should test the validity of this assumption by testing for change in active days or visit per user due to the treatment.

Under those assumptions, the observed outcome of a user can be modelled as [27]:
$$Y_u(t) = R_u(t)\{Z_u \tau_u(t) + c_u(t) + \epsilon\} \quad (1)$$
Table 1 below describes the notations used above. If a user $u$ in the control group is present on day $t$, then its observed outcome would be $c_u(t)$, and if a user $u$ in treatment is present on day t, then the observed outcome would be $\tau_u(t) + c_u(t)$. Note that if a user $u$ is not present on day $t$, then the outcome would be zero, i.e., $Y_u(t) = 0$. When a zero is shown in the analysis, it could be either from an observed zero outcome or from the absence in the data.

**Bias**

We define bias as the difference in the expected treatment effect estimate and the true treatment effect on a user under the model described in equation (1). This follows the standard definition of bias in a parametric model [12].

**Variance and Statistical Power**

We follow the standard definitions for variance and statistical power [20]. Variance is a measure of the uncertainty in our estimate of the treatment effect. Statistical power is the chance we will detect a statistically significant treatment effect when there is indeed a non-zero treatment effect. The higher the variance of a metric, the lower is the statistical power to detect a predetermined treatment effect.

Table 1 Notations used across the paper

| Notation | Explanation |
|---|---|
| $Y_u(t)$ | The observed outcome of user u at day t |
| $R_u(t)$ | Identity function for presence of activity from user $u$ on day $t$ |
| $Z_u$ | Indicator of a user $u$ being assigned to treatment |
| $t_u^0$ | The first time a user $u$ has an activity in the experiment window i.e. $\min\{t|R_u(t) = 1\}$ |
| $c_u(t)$ | Control outcome for user $u$ at day $t$ |
| $\tau_u(t)$ | Treatment effect for user $u$ at day $t$ |
| $k$ | Duration of the experiment (from day 1 to k) |
| $a$ | The last day of admission of new users. It is $k - d$ for bounded and $k$ for open approach |
| $d$ | Duration of observation period (bounded) |
| $\theta \in \{o, b\}$ | Data inclusion approach in analysis – either bounded ($b$) or open ($o$) |
| $\mathbb{I}_{ub}, \mathbb{I}_{uo}$ | Interval of days included in bounded and open analysis for a user $u$ |
| $N_b, N_o$ | Total number of users in bounded and open analysis |
| $T, C$ | Superscripts $T$ and $C$ are used to indicate parameters for treatment and control |

### 2.3 Bounded and Open Approaches

**Open approach:** In open analysis, observations from all users $u \in U$ in all $k$ days are used for estimating the average treatment effect. For any user who is exposed to the experiment within the $k$ days of the experiment, we include all the observations from their first day of presence, i.e., $\mathbb{I}_{uo} = [t_u^0, k]$, (see Figure 1).

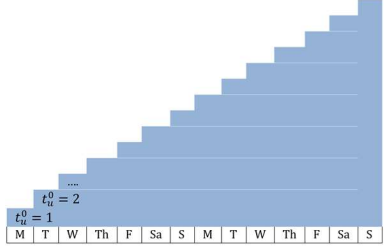

**Figure 1:** Visual representation of the open data inclusion approach.

**Bounded approach:** In bounded analysis, for any user $u \in U$ that is first exposed to the experiment on day $t_u^0 \leq k - d$, we observe the user for the interval of days $\mathbb{I}_{ub} = [t_u^0, t_u^0 + d)$. Observations from any user that is first exposed to the experiment on day $t_u^0 > k - d$ will not be included in the analysis (see Figure 2).

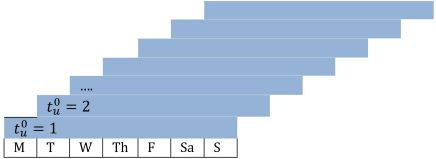

**Figure 2:** Visual representation of the bounded data inclusion approach. The figure represents the admitted cohorts and their bounded observation for one week.

### 2.4 Metric Definition

Many different types of metrics can be computed in analyzing experiments. In this paper, we focus on double average metrics which is the total metric sum per user per active day in the treatment/control group, e.g., average revenue per user per active day.

For bounded ($\theta = b$) and open ($\theta = o$) analysis methods, the first average is computed over the number of active days for a given user observed under either treatment ($z = T$) or control ($z = C$):

$$m_{u\theta}^z = \frac{\sum_{t \in \mathbb{I}_{u\theta}} Y_u(t, z_u = z)}{\sum_{t \in \mathbb{I}_{u\theta}} R_u(t)} \quad (2)$$

Then we calculate sample mean of the metric by taking the average across all users within treatment ($z = T$) and control ($z = C$) sample.

$$\widehat{M}_{u\theta}^z = \frac{1}{N_\theta^z} \sum_{i=1}^{a} \sum_{u:t_u^0=i, z_u=z} m_{u\theta}^z \quad (3)$$

which is an estimate of the population average in expectation

$$\mathbb{E}(\widehat{M}_{u\theta}^z) = \frac{1}{\mathbb{E}[N_\theta]} \sum_{i=1}^{a} \sum_{u:t_u^0=i} m_{u\theta}^z \quad (4)$$

Under potential outcomes framework the treatment effect is estimated by the difference in sample means under treatment and control [14]:

$$\widehat{\Delta}_\theta = \widehat{M}_{u\theta}^T - \widehat{M}_{u\theta}^C \quad (5)$$

In the experimentation frameworks, the variance of the $\hat{\Delta}$ is not directly measurable and is estimated using the pooled sample variances of observed treatment and control samples:

$$\text{Var}(\widehat{\Delta}_\theta) = \frac{\widehat{s_{M_{u\theta}^T}^2}}{N_\theta^T} + \frac{\widehat{s_{M_{u\theta}^C}^2}}{N_\theta^C} \quad (6)$$

and

$$\mathbb{E}[\text{Var}(\widehat{\Delta}_\theta)] \approx \frac{\mathbb{E}\left[\widehat{s_{M_{u\theta}^T}^2}\right]}{\mathbb{E}[N_\theta^T]} + \frac{\mathbb{E}\left[\widehat{s_{M_{u\theta}^C}^2}\right]}{\mathbb{E}[N_\theta^C]} \quad (7)$$

There are other common metric categories like single average metrics and proportion metrics that are out of scope for this paper and will be subject of future research.

## 3 Bias and Variance Tradeoff

In this section we mathematically model the bias and variance estimates for the bounded and open approaches under various assumptions. We encourage the readers to also reference the supplementary material for more details [29].

Let us first analyze the bias of the estimated treatment effect in the simplest case – constant treatment effect. Note that practically this will also include cases where the variation in treatment effect is too small to change a ship decision.

**Result 1:** For a double average metric, if there is a constant treatment effect both bounded and open analysis will yield unbiased estimates of the treatment effect.

$$\mathbb{E}[\widehat{M}_\theta^T - \widehat{M}_\theta^C] = \frac{1}{\mathbb{E}[N_\theta]} \sum_{i=1}^{a} \sum_{u:t_u^0=i} \mathbb{E}[M_{u\theta}^T] - \mathbb{E}[M_{u\theta}^C] \quad (8)$$
$$= \tau$$

We need to make a few more assumptions regarding user behavior to study the bias variance tradeoff between the two approaches. The same assumptions also help us model the bias variance trade off in a more general case of time varying treatment effect.

### 3.1 User Population and Engagement Models

We focus on two types of user population, which are the most common based on our experience at Microsoft – a fixed target population model for online service experiments (e.g. Bing), and an evolving target population models for client experiments (e.g. Edge). For brevity and simplicity, we will fix $k = 14$ and $d = 7$ in the model described by equation (1). This is the most common setting for experiments at Microsoft.

### 3.2.1 Model 1 (Constant treatment effect with additional weekend effect, Fixed target population, Random engagement)

For online service experiments, once a code change for an experiment is deployed to the servers, every user request goes to through a single code base and has the potential to be part of the experiment. Within the code, there may be a lot of code branches that a user may take based on their membership in certain experiments and other user characteristics. Thus, there is a single and fixed target population of users that the service is catering.

In this model, we make the following assumptions on user activity and treatment effect:

- User activity is random with the probability of a user $u$ showing up at day $t$ equals to $p$, i.e.,
$$R_u(t) = Bernoulli(p) \quad for\ t \geq t_u^0 \quad (5)$$
- The user treatment effect is constant with an additional interaction effect ($\tau'$) on weekend. The user control value is a constant with an additional error term which follows a normal distribution,
$$\tau_u(t) = \tau + \tau' I(t = weekend),$$
$$c_u(t) = c_u + \mathcal{N}(0, \sigma^2). \quad (6)$$

The first assumption is a very simple representation of a fixed target population sample in the experiment. In reality, the activity pattern $R_u(t)$ is dependent on past activity and can vary across time and users. The second assumption is a simple representation of a treatment with different effects on weekday and weekends.

**Result 2:** Under Model 1
1. Open analysis has no bias in estimating the treatment effect in presence of a weekend effect. Bounded analysis has a slight bias which in the worst case can underestimate the treatment effect by $0.064\tau'$. Figure 3.a shows the estimated bias in open and bounded analysis for different values of $p$.
2. The bounded analysis leads to an underestimate of the weekend effect if the experiment starts early in the work week (e.g. Monday). Conversely, it will lead to an overestimate of the weekend effect if the experiment starts late in the work week (e.g. Friday).
3. The open analysis method is expected to have a smaller variance. Taking into account the likely values of p, $\tau'$ and $\sigma$ in practice, the variance in the bounded method is more than 50% more than the variance in the open method (as shown in Figure 3.d).

**Proof:**
$$\mathbb{E}[\widehat{\Delta_\theta}] = \mathbb{E}[\widehat{M}_{u\theta}^T] - \mathbb{E}[\widehat{M}_{u\theta}^C] \approx$$
$$\frac{1}{\mathbb{E}[N_\theta]} \sum_{i=1}^{a} \sum_{u:t_u^0=i} \mathbb{E}(M_{u\theta}^T - M_{u\theta}^C) = \tau + \quad (9)$$
$$\frac{1}{\mathbb{E}[N_\theta]} \sum_{i=1}^{a} \sum_{u:t_u^0=i} \tau' \mathbb{E}\left(\frac{\sum_{t \in \mathbb{I}_{u\theta}} R_u(t)(I(t=Weekend))}{\sum_{t \in \mathbb{I}_{u\theta}} R_u(t)}\right)$$

thus, the bias in the estimated treatment effect is
$$\mathbb{E}[\widehat{\Delta_\theta}] - \left(\tau + \frac{2}{7}\tau'\right) =$$
$$\frac{\tau'}{\mathbb{E}[N_\theta]} \sum_{i=1}^{a} \sum_{u:t_u^0=i} \mathbb{E}\left(\frac{\sum_{t \in \mathbb{I}_{u\theta}} R_u(t)(I(t=Weekend))}{\sum_{t \in \mathbb{I}_{u\theta}} R_u(t)}\right) - \frac{2}{7}\tau' \quad (10)$$

To extend this to model 1 for bounded analysis, using equation 10, the bias in the analysis is:
$$\frac{\tau'}{1-(1-p)^7}\left(\sum_{i=1}^{5}(1-p)^{i-1}p\left(\sum_{j=0}^{4}\binom{4}{j}p^j(1-p)^{4-j}\left(\frac{2p(1-p)}{2+j}+\frac{2p^2}{3+j}\right)\right) + \sum_{i=6}^{7}(1-p)^{i-1}p\left(\sum_{j=0}^{5}\binom{5}{j}p^j(1-p)^{5-j}\left(\frac{1-p}{1+j}+\frac{2p}{2+j}\right)\right)\right) - 2/7\ \tau' \quad (11)$$

where $i$ is the first day the user became active, $j$ is the number of active weekdays (other than the first active day of the user). Figure 3.a plots this bias over different values of $p$.

More generally if there are cyclical effects that influence the treatment effect, then bounded analysis has more bias than open analysis. The bounded analysis may underestimate the treatment effect if the cyclic effect occurs at the end of the admission period, or it may overestimate the treatment effect if the cyclic effect occurs at the beginning of the admission period. Intuitively, the bias originates from the difference in the ratio of active days with extra treatment to all active days in the analysis sample and population.

Using similar calculations to analyze the bias in open analysis in a 14-day period with 4 weekend days, equation 10 reduces to

$$\frac{\tau'}{1-(1-p)^{14}} \sum_{i=0}^{10} \binom{10}{i} p^i (1-p)^{10-i} \left(\sum_{j=0}^{4} \binom{4}{j} p^j (1-p)^{4-j} \frac{j}{i+j}\right) - \frac{2}{7}\tau' = 0 \quad (12)$$

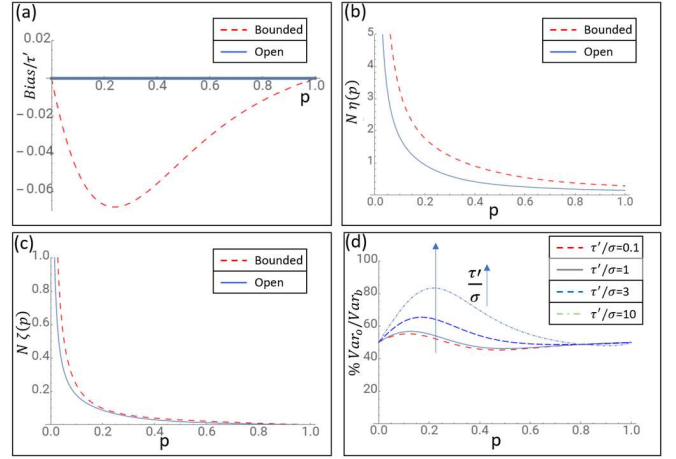

**Figure 3** Bias and variance for the two analysis methods. a) Bias of open and bounded methods under model 1. The absolute bias in open analysis is always smaller than the bias in bounded analysis. b) Coefficient of base activity part of the variance. c) Coefficient of the week-day interaction part of the variance. d) Ratio of the variances of the two methods under different values of $\tau'$, $\sigma$ and $p$. The variance of the open method is more than 50% smaller than that of the bounded.

The variance can also be computed based on breakdown of users on their number of active week and weekend days. The expected value of the sample variance is:

$$\mathbb{E}\left[\widehat{s_{M_{u\theta}^z}^2}\right] = \mathbb{E}\left[\frac{1}{N_\theta}\sum_{i=1}^{a}\sum_{u:t_u^0=i}\left(M_{u\theta}^z - \widehat{M}_{u\theta}^z\right)^2\right]$$
$$\approx \frac{1}{\mathbb{E}[N_\theta]}\sum_{i=1}^{a}\sum_{u:t_u^0=i}\mathbb{E}\left[\left(M_{u\theta}^z - \widehat{M}_{u\theta}^z\right)^2\right] \quad (13)$$

Analysis of variance requires taking a close look at the building blocks of the variance of these estimators, there are three components: a variation in the base user activity level when the user is indeed present on a certain day, ($Var_p$) ,a variation on the week-day interaction term, i.e. the variation of the number of weekends a user is present in the study ($Var_w$). There is also variation in the number of active days for a user ($Var_d$), resulting in a variance is user level metrics. The aggregate effect of these factors will be scaled down by the total number of admitted users – $N_b$ for bounded and $N_o$ for open approach.

Since for the double average metric, the aggregate user level activities are normalized by the number of active days, $Var_d$ does not directly surface in the calculation of the variance and expected variance of these methods would be of the form

$$\mathbb{E}[\text{Var}(\widehat{\Delta}_\theta)] \approx \eta_\theta \sigma^2 + \zeta_\theta \tau'^2 \quad (14)$$

where $\eta_\theta$ is the weight of $\sigma^2$ (variance in the user outcome in control ($C_{u(t)}$)) in the expected variance of estimated treatment effect ($\mathbb{E}[\text{Var}(\widehat{\Delta}_\theta)]$). $\zeta_\theta$ is the weight of $\tau'$ (additional treatment effect on weekend) in $\mathbb{E}[\text{Var}(\widehat{\Delta}_\theta)]$. Both coefficients $\eta_\theta$ and $\zeta_\theta$ are inversely proportional to the number of admitted users.

Since both models have similar $Var_p$ and open analysis admits more user, (i.e. $N_o > N_b$), $\eta_o$ is expected to be smaller than $\eta_b$(Figure 3.b). The coefficient of the week-day interaction term, $\zeta$, has factors moving in opposite direction for the two methods. Open method has higher variation in the number of weekends, $Var_w$. This increase should be contrasted to the increase in the sample size to compare $\zeta_o$ and $\zeta_b$. Such effect is depicted in Figure 3.c. The bounded method always have similar or larger coefficient for the weekday portion of the total variance. The total estimation of variance would be dependent on the magnitude of $\sigma$ and $\tau'$. The result of such analysis is depicted in Figure 3.d where the ratio of the total variances of these method are plotted for different values of $\sigma$, $\tau'$ and $p$. Broadly you can see that the open method has at least 20% smaller variance than the bounded even for unrealistically large time varying effects. Usually, you can expect that most experiments have small average treatment effect and they will have a small weekend effect ($\tau'$). So typical value of $\tau'/\sigma$ can be expected to be small where open analysis performs even better.

Model 1 is based on the fixed target population model. Our analysis of the evolving target population involves a simpler user model, where each user will always be present in the experiment after their first admission.

### 3.2.2 Model 2 (Constant treatment effect with additional weekend effect, Evolving target population, Fixed engagement)

For client experiments, once a code change for an experiment is released in version $v_{next}$, the users who upgrade to version $v_{next}$ will be the ones who will have the potential to be part of the experiment. In this case there are multiple versions of codebase for the same product and different but changing set of users are on different versions. There is usually a growing set of users who will be part of the new version of the product where they may become part of an experiment.

In this model, we make the following assumptions on user activity and treatment effect:

- Users will always be present in the experiment after their first admission
$$R_u(t) = 1 \quad for\ t \geq t_u^0 \quad (15)$$
- Users enter the experiment at a constant rate, $N_G^{i,T} = N_G^{i,C} = N_s$, where $N_s$ is a constant.
- The average treatment effect and users control activity are constant over time with an additional interaction effect ($\tau'$) on weekend, i.e.
$$\tau_u(t) = \tau + \tau' I(t = weekend)\ ,$$
$$c_u(t) = c_u + \mathcal{N}(0, \sigma^2). \quad (16)$$

The first assumption is based on the client-based experiments (e.g., Edge) where users are admitted to the experiment after they update. It is a simplification to assume that users are active every day. In practice we expect users to be active on large number of days. The second assumption is also a simplification of the rate of users updating, but for the simplicity of the model and proof, we assume the population increases at a constant rate.

**Result 3:** Under Model 2,
1. Bounded analysis has no bias in estimating the treatment effect in presence of a weekend effect. Open analysis has a slight bias which can overestimate the treatment effect by $0.19\tau'$ if the experiment starts on Monday. Conversely, it will lead to an underestimate of the weekend effect if the experiment starts late in the work week (e.g. Friday).
2. The bias will continue to decrease if the experiment is run longer ($0.11\tau'$ for 28 days).
3. The variance of bounded analysis is about 20% larger than the variance in the open analysis for the likely values $\tau'$ and $\sigma$ in practice.

Table 2 summarizes the calculated bias and variance under model 2. Because of the assumption of assumption of constant user activity after admission, zero bias is expected for the bounded method as each user is exactly observed for 2 weekends and 5 weekdays. Open approach will have a bias in the estimation of weekend days. The users that enter the experiment close the end, only reflect the treatment associated with those last days. Therefore, the cyclic effect can be over- or under-estimated if it happens on those last days or not. Since the days of user activity are deterministic under this model, the variance in the estimated treatment effect is composed of the variance in the inherent user behavior ($C_u$) that is scaled down by the total admitted users ($aN_s$) due to averaging. Therefore, a much smaller variance is expected for the open method due to larger number of admitted users.

**Table 2** Bias and variance of open and bounded approaches under model 2

| Metric | | bounded | open |
|---|---|---|---|
| Double average | $bias$ | 0 | $0.19\tau'$ |
| | $var$ | $\frac{0.041}{N_S}\sigma^2$ | $\frac{0.033}{N_S}\sigma^2 + \frac{0.004}{N_S}\tau'^2$ |

## 4 Empirical evaluation using simulations

In this section, we use simulations with real-world data to evaluate the accuracy and power of the open and bounded approaches. These simulations run on a sample of a key Bing metric (fixed target population) from 13 million users, and a sample of a key Microsoft Edge browser metric (evolving sample) from 5 million devices that elected to share their data. These simulations accurately reflect real-world variations in metrics while allowing us to insert known treatment effect for evaluation (i.e. we know the actual treatment effect).

Recall from equation (1), the metric of interest is modelled as:
$$Y_u(t) = R_u(t)\{Z_u \tau_u(t) + c_u(t) + \epsilon\} \quad (17)$$

In the simulation, we retain the user model from the real-world data so that the user activity pattern, $R_u(t)$, and the metric of interest in control group, $c_u(t)$, is not influenced by our simulation. We only simulate the treatment assignment $Z_u$ and the treatment effect $\tau_u(t)$. The metric we are simulating on is a double average, defined by equation (2).

To generate the data, we randomly assigned users to treatment and control groups. Then we artificially injected a treatment effect ($\tau$) for those users in treatment. We will cover two simulations in the sections below: time-invariant treatment effect and time-variant treatment effects. We estimated the average treatment effect ($\widehat{\Delta}$) as the differences in averages between treatment and control using both open and bounded analysis. Finally, we tested for statistical significance of the estimated effect. We used increasingly larger sample sizes, and at each step we repeated the process 500 times to obtain the p-values, as well as the $5^{th}$, $50^{th}$ (median), and $95^{th}$ percentiles of the treatment effect.

We plot the probability of being statistically significant (power) for each sample size for both bounded and open approaches. We also plot the median point estimate for the difference (with 5 and 95 percentiles as bootstrap confidence intervals), which assesses accuracy.

In both plots, the x-axis is sample percentage from the total data set. However, since the open and bounded approaches have different rules for which users to include, the actual number of users used in computations differ. The open approach includes more users, since the bounded approach discards users who appear later in the experiment. This helps reduce variance in the open method. Though, users that are active later in the experiment can potentially behave quite differently from the users who are active in the experiment early thus introducing noise and increasing variance in the analysis. Overall, as the sample percentage increases, we expect to both approaches to increase in the probability of detecting the known change. It is the difference in power between the two approaches that is of interest.

### 4.1 Time-Invariant Treatment Effect

In this simulation we introduce a fixed treatment effect $\tau$ for all users on all days i.e., $\tau_u(t) = \tau$ for all users $u$ and time $t$. We choose $\tau$ such that there is a 1% lift in treatment on average.

An unbiased (accurate) estimate should show the same constant percentage lift, i.e., the point estimate should equal 1%. In addition, the more powerful approach should have a tighter confidence interval.

As argued in the previous sections, we expect the point estimates for the treatment effect for both open and bounded approaches to be the same.

Power and point estimate plots for the bounded and open approaches data are in Figure 4.1 and Figure 4.2. Both approaches produce similarly accurate point estimates of the treatment effect. However, with the same percentage of available data, the open approach consistently has more power, having a higher probability of detecting the difference and tighter confidence intervals. This shows that benefits of having more data (in the open approach) dominates the increased variance in the data itself

We also note that the Microsoft Edge and Bing metrics have very high variance, with each having values more than twice the mean. These differences in power would be even more pronounced for metrics with lower variances.

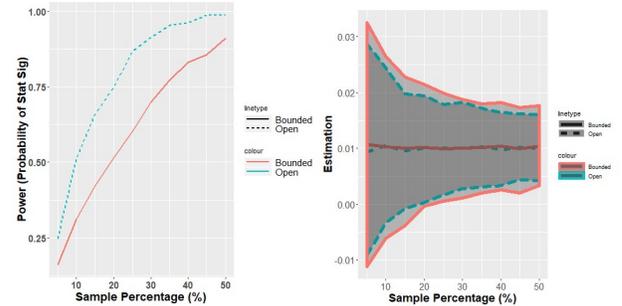

**Figure 4.1** Power and point estimate plots for Bing data with fixed treatment effect

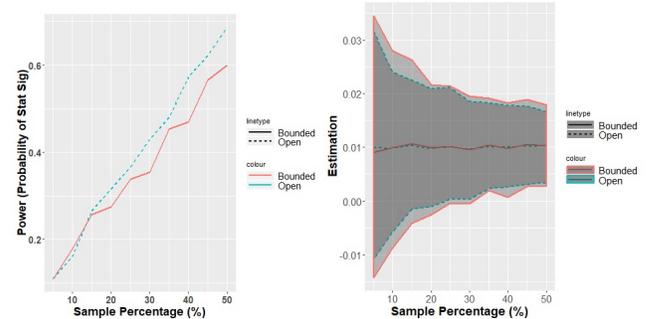

**Figure 4.2** Power plot and point estimate plots for Edge data with fixed treatment effect

### 4.2 Time-varying Treatment Effect

When the treatment effect is time-varying, open and bounded approaches can have differences point estimates (i.e. differences in accuracy). We examine the difference by introducing different weekday and weekend treatment effects. In our second set of simulations we inject a large treatment effect, $\tau' = 10\tau$, for weekends. This is in addition to the treatment effect $\tau$ that we introduced in the section before. On average, the treatment observations will have a treatment effect of $\tau + \tau'$ (11%) on weekend and $\tau$ (1%) on weekdays.

As shown in the section above, the overall treatment effect can be obtained by:

$$\mathbb{E}[\hat{\Delta}] = \tau + \tau'\gamma, \quad (18)$$

where $\gamma = \frac{1}{N}\sum_u \frac{\sum_t R_u(t=weekend)}{\sum_t R_u(t)}$. Intuitively, $\gamma$ is the average of the ratio between a user's weekend active days and overall active days.

Power and point estimate plots for the bounded and open approaches are in Figure 4.3 and Figure 4.4. Since the overall treatment effect is large, both approaches are better able to detect differences (i.e. the power curve is steeper), though the open approach is still better. The open approach also provides tighter confidence intervals (same as before). Bing data (fixed target population model) has minimal differences in point estimates of open and bounded methods. This is consistent with the theoretical model estimates in the previous section.

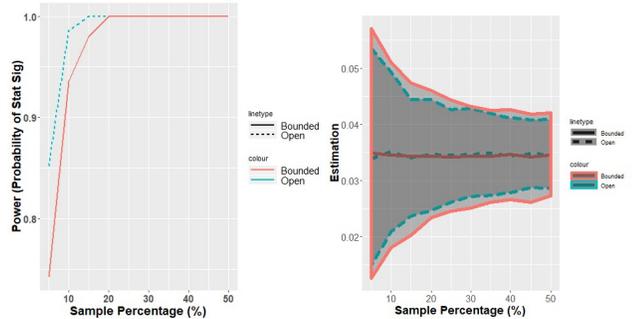

**Figure 4.3** Power and point estimate plot for Bing data with time-variant treatment effect

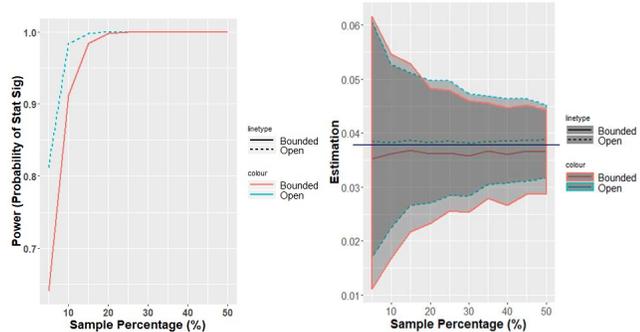

**Figure 4.4** Power and point estimate plot for Edge data with time-variant treatment effect

Edge data (evolving target population model) also does not show a significant difference in the estimates. The confidence intervals overlap by a lot and there is some difference in the point estimates. The open approach's point estimate is 3.85% while bounded approach's point estimate is 3.65% (at 50% sample); the long-term unbiased estimate of the effect (based on the unbiased estimate for $\tau + \tau'\gamma$) is 3.8%. The point estimate of open approach is closer to the long-term estimate than the bounded approach. This may appear at odds with the results in Table 2. We realize that model 2 is very simple and Edge data may have much more complex underlying patterns for user activity. It is hard to read too much into the difference between the point estimates as the confidence intervals overlap by a lot. The point estimate in bounded approach can vary more due to larger variance than in the open approach. The key takeaway from this simulation is that the estimates might vary between open and closed approach for evolving target population but the difference between the two is very small to be statistically significant.

## 5  Empirical study of real-world experiments

In this section we further compare bounded and open approaches using a real-world large-scale digital randomized controlled experiment. Unlike simulations in the previous section, we do not know the actual treatment effects. The effects may not be the same for all users and may be different across different days (in addition to weekday/weekends effects). In addition, extraneous factors (e.g. holidays days) may also interact with the treatment. Therefore, this evaluation will provide us understanding of the practical differences between the approaches under real-world conditions.

We examine a Microsoft Edge web-browser experiment to assess the impact of adding translation functionality in the browser. When the user is on a page with content in a different language compared to operating system, the new functionality shows the user the option to translate the content of the page. This experiment ran for a month (28 days) in early 2018 with over 25 million users world-wide.

Part way through the experiment, an issue occurred that caused a known time varying effect.

Figure 5.1 shows the power and point estimate plots for the double average of browser problem reports metric for the experiment. Similar to simulations, we used increasingly larger sample sizes, and at each step we repeated the process 500 times to obtain the p-values, as well as the $5^{th}$, $50^{th}$ (median), and $95^{th}$ percentiles of the treatment effect. As with simulations, the open approach has power (steeper power curves and tighter confidence intervals) in detecting the increase in the metric compared to the bounded approach. At 100% sample we get just one-point estimate which is likely the best point estimate of the treatment effect. The point estimate of treatment effect for open approach is slightly higher than that of the bounded approach. This is consistent with results in both mathematical derivations and simulations. This is likely attributable to the evolving target population.

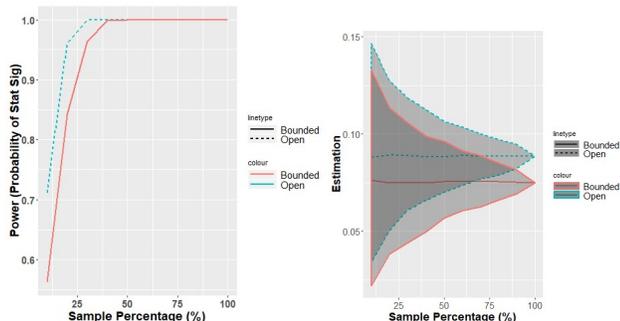

**Figure 5.1** Power and point estimate plot for the Edge experiment

We have analyzed dozens of metrics of different types (single average, double average, proportion) in a large number of experiments run over the last few years with both open and bounded analysis. In our experience, the open analysis had higher statistical power. The point estimates of treatment effect from both open and bounded approaches were practically similar - bias introduced by open analysis did not alter decisions.

## 6  Conclusion

In this paper, we have studied the tradeoff between accuracy (bias) and statistical power (variance) for open and bounded approaches for data inclusion in the analysis of randomized controlled experiments, under various assumptions about user populations – fixed target population for service experiments and evolving target population for client experiments. We have further evaluated the approaches with simulations using real-world data and an actual Edge experiment. Our conclusion is that the open approach is preferable.

For time-invariant homogeneous treatment effect, we show that both approaches will produce an estimate of the treatment effect that has no bias in the case of double average metrics. This result also applies in cases where the heterogeneous treatment effect is small. We find that the open approach can have biased estimates of treatment effect in the presence of time-variant treatment effects (e.g. weekday /weekend effects) in evolving user populations. The bounded approach can help correct the bias; however, the bounded approach sacrifices statistical power.

In practice the bias is small and, as we have experienced, the ship decision generally did not change due to small bias. Therefore, in general, we recommend practitioners use the open approach to have the ability to detect small changes (i.e. high statistical power). We also recommend practitioners to segment experiment results by date to detect cases where the time varying effect is very large and it would become a factor in decision making.

**Future Work**

The study of difference between open and bounded approach for data inclusion can be investigated for a large variety of cases. This paper included only one type of metrics – double average. We plan to study other types of metrics (e.g. single average, proportions and distributions) that may have different characteristics. Also, different assumptions about user populations (e.g. heavy and light users) and treatment effects (e.g. learning effects) are of interest to model and examine empirically.

## ACKNOWLEDGMENTS

We are grateful to our colleagues in Microsoft for who helped run these experiments and helped shape this study. We would also like to thank Pavel Dmitriev and Yu Wang for their help in early discussions and framing for this problem, and Widad Machmouchi and Benjamin Arai for their valuable suggestions to this paper.

## S1: Intuition about the root of the bias under model 1

Intuitively, the bias originates from the difference in the ratio of active weekend days to all active days in the analysis sample and population. Let's study a very simple model to develop this intuition further for bounded analysis. Table S3 illustrates the simplest model to study time varying treatment effect for bounded analysis: an experiment with total length of 4 days where the even days have an additional treatment effect $\tau'$. The bounded analysis has an admission window of 2 days ($a = 2$) and observation window of 2 days (d=2). Each user's active days list is shown with 4 characters, 1 for active, 0 for absent and x for 'don't care' which is shown in column 1. The desired value for the ratio of number of even days to all days is 0.5 but different sets of users show values between 0 and 1.

**Table S3:** Illustration of cause of bias in the bounded approach in a simple experiment with even day effect and an admission period and observation period of 2 days.

| Active days of the user | Probability | #Active even days | #Active days logged | Expected ratio |
|---|---|---|---|---|
| 10xx | $(1-p)p$ | 0 | 1 | 0 |
| 11xx | $p^2$ | 1 | 2 | $0.5p^2$ |
| 010x | $(1-p)^2 p$ | 1 | 1 | $(1-p)^2 p$ |
| 011x | $p^2(1-p)$ | 1 | 2 | $0.5p^2(1-p)$ |
| 0010 | $(1-p)^2 p$ | 0 | 1 | 0 |
| 0011 | $(1-p)^2 p^2$ | 1 | 2 | $0.5(1-p)^2 p^2$ |

The expected value of the average number of active even days is now $\frac{0.5p^2 + p(1-p)^2 + 0.5p^2(1-p) + 0.5(p^2)(1-p)^2}{1-(1-p)^4}$ which is different from the theoretical value of 0.5 and varies between 0.25 and 0.5 depending on the value of p.

Using the open method, the desired value for the ratio of number of even active days to all active days is still 0.5. Table S2 shows the breakdown of possible outcomes of user activity and its probability.

**Table S2:** Illustration of expected observation for the open approach in a simple experiment with even day effect and duration of 4 days.

| Active days of the user | Probability | #Active even days | #Active days logged |
|---|---|---|---|
| 1000, 0010 | $2(1-p)^3 p$ | 0 | 1 |
| 0100, 0001 | $2(1-p)^3 p$ | 1 | 1 |
| 1001, 1100, 0110, 0011 | $4(1-p)^2 p^2$ | 1 | 2 |
| 1010 | $(1-p)^2 p^2$ | 0 | 2 |
| 0101 | $(1-p)^2 p^2$ | 2 | 2 |
| 1011, 1110 | $2(1-p)p^3$ | 1 | 3 |
| 1101, 0111 | $2(1-p)p^3$ | 2 | 3 |
| 1111 | $p^4$ | 2 | 4 |

Although different users observe different number of even days, the expected value for the average of these users is

$\left(2(1-p)^3 p \left(\frac{1}{1}\right) + 4(1-p)^2 p^2 \left(\frac{1}{2}\right) + (1-p)^2 p^2 \left(\frac{2}{2}\right) + 2(1-p)p^3 \left(\frac{1}{3}\right) + 2(1-p)p^3 \left(\frac{2}{3}\right) + p^4 \left(\frac{2}{4}\right)\right) / (1 - (1-p)^4) = \frac{1}{2}$

which shows no bias for the open method under this scenario.

## S2: Proof of Results Under Model 1

Let $G_i$ be the set of users with first day as $i$, then the size of each cohort, i.e., the number of users entering the experiment on each day is $N_G^i$ with the expected value of

$$\mathbb{E}[N_G^i] = N(1-p)^{i-1} p, \quad (S1)$$

and the total number of admitted users is:

$$\mathbb{E}[N_b] = \sum_{i=1}^{a} \mathbb{E}[N_G^i] = N(1 - (1-p)^a). \quad (S2)$$

Let's start by the bounded methods. There are several possibilities for $m_{ub}^z$ based on the user's number of active weekdays and weekends, let $w$ be the iterator over these possible outcomes, and $N_G^w$ be the number users with that number of weekdays and weekends ($w$ iterates from 0 weekday, 0 weekend to 5 weekday, 2 weekends):

$$\mathbb{E}[\hat{M}_{ub}^z] = \frac{1}{\mathbb{E}[N_b]} \sum_{i=1}^{7} \sum_{u:t_u^0=i} m_{ub}^z$$

$$= \frac{1}{N(1-(1-p)^7)} \sum_{i=1}^{7} \sum_w \mathbb{E}[N^w] \mathbb{E}[m_{ub}^{z,w}] =$$

$$\frac{1}{1-(1-p)^7} \sum_{i=1}^{5} (1-p)^{i-1} p \left( \sum_{j=0}^{4} \binom{4}{j} p^j (1-p)^{4-j} \left( (1-p)^2 \left( C_u + Z(\tau + \frac{0\tau'}{1+j}) \right) + 2p(1-p) \left( C_u + Z(\tau + \frac{1\tau'}{2+j}) \right) + p^2 \left( C_u + Z(\tau + \frac{2\tau'}{3+j}) \right) \right) \right) + $$

$$\sum_{i=6}^{7} (1-p)^{i-1} p \left( \sum_{j=0}^{5} \binom{5}{j} p^j (1-p)^{5-j} \left( (1-p) \left( C_u + Z(\tau + \frac{1\tau'}{1+j}) \right) + p \left( C_u + Z(\tau + \frac{2\tau'}{2+j}) \right) \right) \right) \quad (S3)$$

where $i$ is the first day the user became active, $j$ is the number of active weekdays (other than the first active day of the user). The bold segments are the expected value of the metric for that subgroup of users. The expected value of the treatment effect, $\mathbb{E}[\hat{\Delta}_b]$, then can easily be calculated.

$$\mathbb{E}[\hat{\Delta}_b] = \mathbb{E}[\hat{M}_{ub}^T] - \mathbb{E}[\hat{M}_{ub}^C] = \tau + \rho_b(p) \tau' \quad (S4)$$

where $\rho_b$ is a polynomial fraction. The expected bias of this method is the different of this value and $\tau + \frac{2}{7} \tau'$ for double average class of metrics as is shown in Figure 3.a.

Similar breakdown of users based on the number of active weekends and weekdays can be done for the open analysis method. There is no need to use the concept of admission here. Each user can be active for up to 10 weekdays and 4 weekend days, let $N_G^{i,j}$ be the number of users in each group:

$$\mathbb{E}[\widehat{M}_{uo}^z] = \frac{1}{\mathbb{E}[N_o]} \sum_{i=0}^{10} \sum_{j=0}^{4} \mathbb{E}[N^{i,j}]\mathbb{E}[m_{uo}^{z,i,j}] =$$
$$\frac{1}{1-(1-p)^{14}} \sum_{i=0}^{10} \sum_{\substack{j=0 \\ j \neq 0 \text{ if } i=0}}^{4} \binom{10}{i} p^i (1- \quad \text{(S5)}$$
$$p)^{10-i} \binom{4}{j} p^j (1-p)^{4-j} \left(C_u + Z\left(\tau + \frac{j\tau'}{i+j}\right)\right)$$

where $i$ and $j$ count the active week and weekend days. Again, the bold segment is the expected value of the metric for that subgroup of users. This equation can be simplified, and it is shown that:

$$\mathbb{E}[\widehat{\Delta}_o] = \mathbb{E}[\widehat{M}_{uo}^T] - \mathbb{E}[\widehat{M}_{uo}^C] = \tau + \frac{2}{7}\tau' \quad \text{(S6)}$$

The expected value of the sample variance can similarly be calculated:

$$\mathbb{E}\left[\widehat{s_{M_{u\theta}^z}^2}\right] = \mathbb{E}\left[\frac{1}{N_\theta} \sum_{i=1}^{a} \sum_{u:t_u^0=i} (M_{u\theta}^z - \widehat{M}_{ub}^z)^2\right] \approx$$
$$\frac{1}{\mathbb{E}[N_\theta]} \sum_{i=1}^{a} \sum_{u:t_u^0=i} \mathbb{E}\left[(M_{u\theta}^z - \widehat{M}_{ub}^z)^2\right] \quad \text{(S7)}$$

For the bounded method, this equation reduces to:

$$\mathbb{E}\left[\widehat{s_{M_{ub}^z}^2}\right] \approx \frac{1}{\mathbb{E}[N_b]} \sum_{i=1}^{a} \sum_{u:t_u^0=i} \mathbb{E}\left[(M_{u\theta}^z - \overline{M_{u\theta}^z})^2\right] = \frac{1}{N(1-(1-p)^7)}$$
$$\sum_{i=1}^{5} (1-p)^{i-1} p \sum_{j=0}^{4} \binom{4}{j} p^j (1-p)^{4-j} \left((1-p)^2 \mathbb{E}\left[\left(\frac{\Sigma^{1+j}\varepsilon}{1+j} + Z\left(\frac{0\tau'}{1+j} - \rho_b(p)\tau'\right)\right)^2\right] + 2p(1-p) \mathbb{E}\left[\left(\frac{\Sigma^{2+j}\varepsilon}{2+j} + Z\left(\frac{1\tau'}{2+j} - \rho_b(p)\tau'\right)\right)^2\right] + p^2 \mathbb{E}\left[\left(\frac{\Sigma^{3+j}\varepsilon}{3+j} + Z\left(\frac{2\tau'}{3+j} - \rho_b(p)\tau'\right)\right)^2\right]\right) +$$
$$\sum_{i=6}^{7}(1-p)^{i-1}p \sum_{j=0}^{5} \binom{5}{j} p^j (1-p)^{5-j} \left((1-p)\mathbb{E}\left[\left(\frac{\Sigma^{1+j}\varepsilon}{1+j} + Z\left(\frac{1\tau'}{1+j} - \rho_b(p)\tau'\right)\right)^2\right] + p\mathbb{E}\left[\left(\frac{\Sigma^{2+j}\varepsilon}{2+j} + Z\left(\frac{2\tau'}{2+j} - \rho_b(p)\tau'\right)\right)^2\right]\right)$$

$$\mathbb{E}[\text{Var}(\widehat{\Delta}_b)] \approx \frac{\mathbb{E}\left[\widehat{s_{M_{ub}^T}^2}\right]}{\mathbb{E}[N_b^T]} + \frac{\mathbb{E}\left[\widehat{s_{M_{ub}^C}^2}\right]}{\mathbb{E}[N_b^C]} \quad \text{(S8)}$$
$$= \eta_b(p)\sigma^2 + \zeta_b(p)\tau'^2$$

where $\eta_b(p)$ and $\zeta_b(p)$ is depicted in **Error! Reference source not found.**.

The expected value of the sample variance for the open method can be calculated using the same breakdown we used for the bias calculation:

$$\mathbb{E}\left[\widehat{s_{M_{uo}^z}^2}\right] \approx$$
$$\frac{1}{\mathbb{E}[N_o]} \sum_{i=0 \text{ to } 10} \sum_{j=0 \text{ to } 4} \mathbb{E}[N_G^{i,j}]\mathbb{E}\left[\left(m_{ub}^{z,i,j} - C_u - Z\left(\tau + \frac{2}{7}\tau'\right)\right)^2\right] =$$
$$\frac{1}{1-(1-p)^{14}} \sum_{i=0 \text{ to }} \sum_{j=0 \text{ to } 4} \binom{10}{i} p^i (1- \quad \text{(S9)}$$
$$p)^{10-i} \binom{4}{j} p^j (1-p)^{4-j} \mathbb{E}\left[\frac{\Sigma^{i+j}\varepsilon}{i+j} + Z\left(\frac{j\tau'}{i+j} - \frac{2}{7}\right)\right]^2$$

$$\mathbb{E}[\text{Var}(\widehat{\Delta}_o)] \approx \frac{\mathbb{E}\left[\widehat{s_{M_{uo}^T}^2}\right]}{\mathbb{E}[N_o^T]} + \frac{\mathbb{E}\left[\widehat{s_{M_{uo}^C}^2}\right]}{\mathbb{E}[N_o^C]} \quad \text{(S10)}$$
$$= \eta_o(p)\sigma^2 + \zeta_o(p)\tau'^2$$

where $\zeta_o(p)$ and $\gamma_o(p)$ is depicted in **Error! Reference source not found.**.

## S3: Proof of Results Under Model 2

We first show that the bounded observation always provides an unbiased estimator under model 2 if observation window is one week.

$$\widehat{\Delta}_b = \widehat{M}_{ub}^T - \widehat{M}_{ub}^C \quad \text{(S11)}$$

$$\widehat{M}_{ub}^z = \frac{1}{N_b^z} \sum_{i=1}^{7} \sum_{u \in G^i} \frac{\sum_{t=i}^{i+d-1} R_u(t)(C_u + Z_u \tau_u(t) + \mathcal{N}(0,\sigma^2))}{\sum_{t=i}^{i+d-1} R_u(t)} \quad \text{(S12)}$$

$$\mathbb{E}[\widehat{M}_{ub}^z] = \frac{1}{N_b^T} \sum_{i=1}^{7} \sum_{u \in G_i} \left(C_u + Z_u\left(\tau + \frac{2}{7}\tau'\right)\right)$$
$$= \frac{1}{N_b^T} \sum_{i=1}^{7} \left(N_s \left(C_u + Z_u\left(\tau + \frac{2}{7}\tau'\right)\right)\right)$$
$$= \frac{1}{7N_s} \left(7N_s\left(C_u + Z_u\left(\tau + \frac{2}{7}\tau'\right)\right)\right) \quad \text{(S13)}$$
$$= C_u + Z_u\left(\tau + \frac{2}{7}\tau'\right)$$

The variance term can similarly be calculated:

$$\widehat{\sigma_{M_{ub}^z}^2} = \frac{1}{N_b^z} \sum_{i=1}^{7} \sum_{u \in G^i} \left(\frac{\sum_{t=i}^{i+6} R_u(t)(C_u + Z_u \tau_u(t) + \mathcal{N}(0,\sigma^2))}{\sum_{t=i}^{i+d-1} R_u(t)} - C_u - Z_u\left(\tau + \frac{2}{7}\tau'\right)\right)^2$$

$$\mathbb{E}\left[\widehat{\sigma_{M_{ub}^z}^2}\right] =$$
$$\frac{1}{\mathbb{E}[N_b^z]} \sum_{i=1}^{7} N_s \mathbb{E}\left[\left(\frac{\sum_{t=i}^{i+6} R_u(t)(C_u + Z_u \tau_u(t) + \mathcal{N}(0,\sigma^2))}{\sum_{t=i}^{i+d-1} R_u(t)} - C_u - Z_u\left(\tau + \frac{2}{7}\tau'\right)\right)^2\right] = \frac{1}{\mathbb{E}[N_b^z]} \sum_{i=1}^{7} N_s \mathbb{E}\left[\left(\frac{\sum_{t=i}^{i+6} \mathcal{N}(0,\sigma^2)}{7}\right)^2\right] = \frac{\sigma^2}{7} \quad \text{(S14)}$$

therefore,

$$\mathbb{E}[\widehat{\Delta}_b] = \tau + \frac{2}{7}\tau' \text{ and } \mathbb{E}[\text{Var}(\widehat{\Delta}_b)] = \frac{2}{49N_s}\sigma^2 \quad \text{(S14)}$$

For open analysis method:

$$\mathbb{E}[\widehat{M}_{uo}^z] = \frac{1}{N_o^z} \sum_{i=1}^{14} \mathbb{E} \sum_{u \in G_i} \frac{\sum_{t=i}^{14} R_u(t)(C_u + Z_u \tau_u(t) + \mathcal{N}(0,\sigma^2))}{\sum_{t=i}^{14} R_u(t)}$$
$$= \frac{1}{N_o^T} \sum_{i=1}^{14} N_s \left(C_u + Z_u\left(\tau + \left(\frac{n_{weekend}(i)}{15-i}\right)\tau'\right)\right)$$
$$\cong \frac{1}{14N_s} N_s(14C_u + Z_u(14\tau + 6.7\tau'))$$
$$= C_u + Z_u\left(\tau + \frac{6.7}{14}\tau'\right)$$

The variance term can similarly be calculated:

$$\widehat{\sigma_{M_{uo}^z}^2} = \frac{1}{N_o^z} \sum_{i=1}^{14} \sum_{u \in G^i} \left(\frac{\sum_{t=i}^{14} R_u(t)(C_u + Z_u \tau_u(t) + \mathcal{N}(0,\sigma^2))}{\sum_{t=i}^{14} R_u(t)} - C_u - Z_u\left(\tau + \frac{6.7}{14}\tau'\right)\right)^2$$

$$\mathbb{E}\left[\widehat{\sigma_{M_{uo}^z}^2}\right] =$$
$$\frac{1}{\mathbb{E}[N_o^z]} \sum_{i=1}^{14} N_s \mathbb{E}\left[\left(\frac{\sum_{t=i}^{i+6} R_u(t)(C_u + Z_u \tau_u(t) + \mathcal{N}(0,\sigma^2))}{\sum_{t=i}^{i+d-1} R_u(t)} - C_u - Z_u\left(\tau + \frac{6.7}{14}\tau'\right)\right)^2\right] \quad \text{(S15)}$$
$$= \frac{1}{\mathbb{E}[N_o^z]} \sum_{i=1}^{14} N_s \mathbb{E}\left[\left(\frac{\sum_{t=i}^{14} \mathcal{N}(0,\sigma^2)}{15-i} + \left(\frac{n_{weekend}(i)}{15-i} - \frac{6.7}{14}\right)\tau'\right)^2\right] = \frac{1}{\mathbb{E}[N_o^z]} \sum_{i=1}^{14} N_s \mathbb{E}\left[\left(\frac{\sum_{t=i}^{14} \mathcal{N}(0,\sigma^2)}{15-i}\right)^2\right] + \mathbb{E}\left[\left(\left(\frac{n_{weekend}(i)}{15-i} - \frac{6.7}{14}\right)\tau'\right)^2\right] = \frac{3.25}{14}\sigma^2 + Z_u\left(\frac{0.76}{14}\tau'^2\right)$$

$$\mathbb{E}[\widehat{\Delta_o}] = \tau + \frac{6.7}{14}\tau' \quad \text{and} \quad \mathbb{E}[Var(\widehat{\Delta_o})] = \frac{6.5}{14^2 N_s}\sigma^2 + \frac{0.76}{14^2 N_s}\tau'^2 \tag{S16}$$